\documentclass[aps,pra,reprint,superscriptaddress]{revtex4-1}

\usepackage{graphicx}
\usepackage{bm}
\usepackage{bbold}

\usepackage{multirow}

\newcommand{\red}[1]{{#1}}

\renewcommand{\vec}[1]{{\mathbf #1}}

\begin{document}

\title{Search for Anderson localization of light by cold atoms in a static electric field}

\author{S.E. Skipetrov}
\email[]{Sergey.Skipetrov@lpmmc.cnrs.fr}
\affiliation{Univ. Grenoble Alpes, CNRS, LPMMC, 38000 Grenoble, France}

\author{I.M. Sokolov}
\email[]{ims@is12093.spb.edu}
\affiliation{Department of Theoretical Physics, Peter the Great St. Petersburg Polytechnic University, 195251 St. Petersburg, Russia}

\date{\today}

\begin{abstract}
We explore the potential of a static electric field to induce Anderson localization of light in a large
\red{three-dimensional (3D)}
cloud of randomly distributed, immobile atoms with a nondegenerate ground state (total angular momentum $J_g = 0$) and a three-fold degenerate excited state ($J_e = 1$). We study both the spatial structure of quasimodes of the atomic cloud and the scaling of the Thouless number with the size of the cloud. Our results indicate that unlike the static magnetic field, the electric field does not induce Anderson localization of light by atoms. We explain this conclusion by the incomplete removal of degeneracy of the excited atomic state by the field and the relatively strong residual dipole-dipole coupling between atoms which is weaker than in the absence of external fields but stronger than in the presence of a static magnetic field.
\red{A joint analysis of these results together with our previous results concerning Anderson localization of scalar waves and light suggests the existence of a critical strength of dipole-dipole interactions that should not be surpassed for Anderson localization to be possible in 3D.
}
\end{abstract}

\maketitle

\section{Introduction}
\label{sec:intro}

Anderson localization of light has been a topic of active research for more than 30 years \cite{john84,anderson85,john91,beek12,segev13,sperling16}. By analogy with electrons in disordered conductors \cite{anderson58}, John \cite{john84} and Anderson \cite{anderson85} independently proposed that optical modes can become exponentially localized in space and light propagation blocked in a sufficiently strongly scattering, disordered dielectric medium (a ``white paint''). This prediction has been experimentally verified in low-dimensional systems \cite{berry97,chabanov00,schwartz07}, but the case of three-dimensional (3D) disorder turned out to be hard to deal with \cite{skip16}.

Cold atoms might represent an alternative to dielectric samples in view of the possible experimental observation of Anderson localization of light because they allow for achieving strong scattering \cite{kaiser99,kaiser09}. However, recent results indicate that the coupling between neighboring atoms by the longitudinal electromagnetic field via the dipole-dipole interaction precludes Anderson localization \cite{skip14,bellando14}. An external magnetic field suppresses this coupling and makes localization of light possible, but strong fields are required \cite{skip15,skip16pra,skip18}.

The strength of dipole-dipole interactions is intimately related with the degeneracy of atomic energy levels. An external magnetic field lifts this degeneracy due to the Zeeman effect but other mechanisms can be envisaged as well. In this work, we explore the possibility of using a static external electric field that affects atomic levels via the Stark effect. Indeed, one might expect the Stark effect to induce Anderson localization, similarly to the Zeeman effect \cite{skip15,skip18}. This would be good news for experiments where reaching strong and spatially uniform electric fields may be easier than creating their magnetic counterparts. However, the Stark effect does not lift the degeneracy of atomic levels completely, in contrast to the Zeeman effect. A three-fold degenerate atomic level with a total angular momentum $J = 1$ (magnetic quantum number $m = 0, \pm 1$), for example, is split into a nondegenerate level $\left| J = 1, m = 0 \right>$ and a two-fold degenerate level $\left| J = 1, m = \pm 1 \right>$ by an external electric field, whereas the magnetic field creates three nondegenerate states corresponding to $m = 0, \pm 1$. This subtle difference stems from the independence of the Stark shift from the sign of $m$ and turns out to be crucial for Anderson localization. We show below that despite the partial removal of atomic level degeneracy by en external electric field, it does not induce Anderson localization of light in a random ensemble of identical two-level atoms.

\section{The model}
\label{model}

\begin{figure}
\hspace*{-1cm}
\includegraphics[width=1.2\columnwidth]{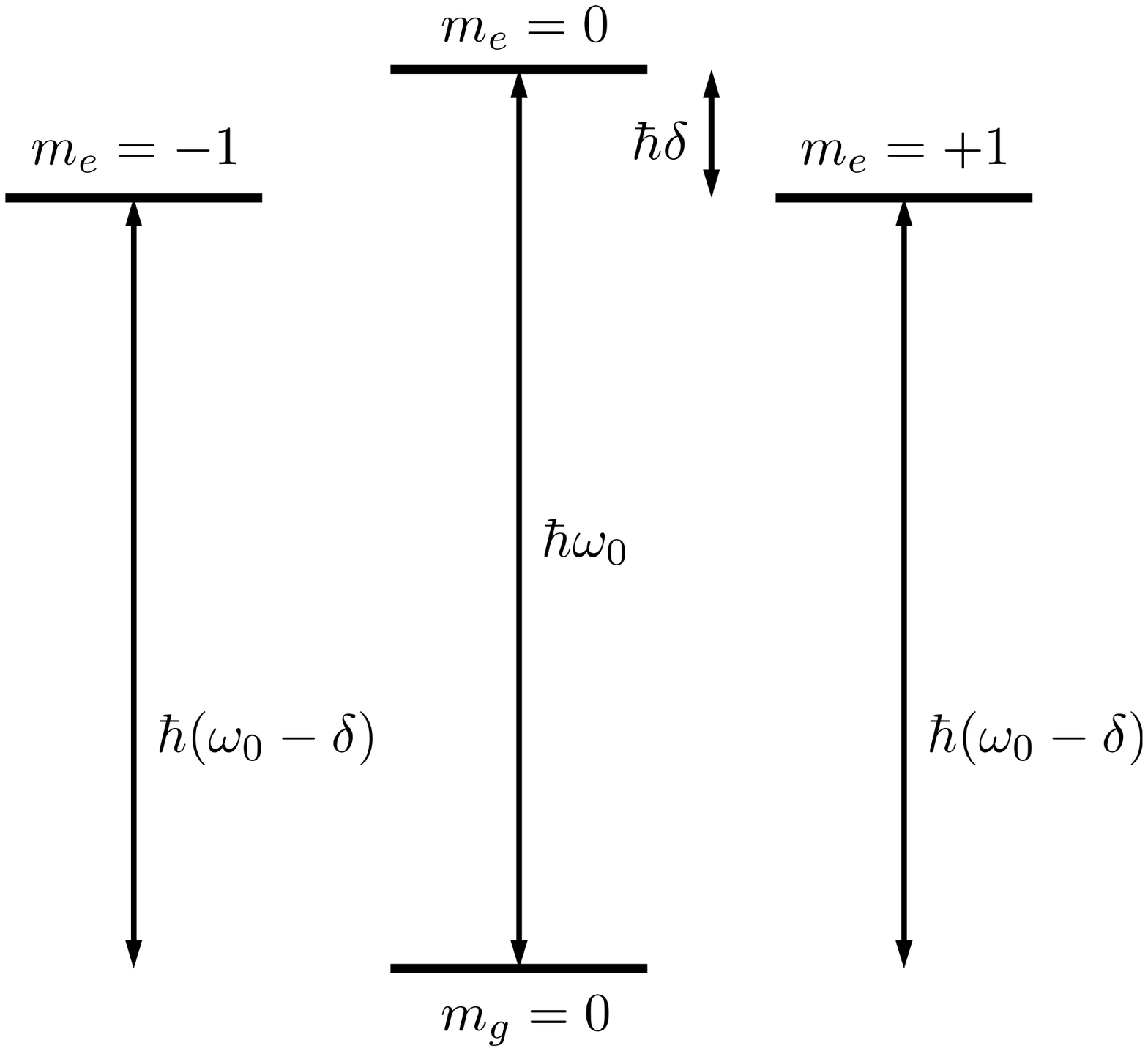}
\vspace*{-1.3cm}
\caption{Schematic energy diagram of a two-level atom in a static electric field.
Horizontal solid lines show positions of energy levels. Level shifts with respect to their positions in the absence of the field are, in the general case, complicated functions of the field strength
but the only important parameters for our analysis are the new resonant frequency $\omega_0$ and the energy difference $\hbar \delta$ between the excited levels corresponding to $m_e = 0$ and $m_e = \pm 1$.}
\label{figlevels}
\end{figure}

A useful approximation to the Hamiltonian of $N$ immobile two-level atoms (ground state $\left| J_g = 0 \right>$, excited state $\left| J_e = 1 \right>$) coupled to the free electromagnetic field and subjected to a spatially uniform, static electric field can be derived by assuming that the static field induces shifts of atomic levels without modifying their lifetimes (the Stark effect) \cite{friedrich90,sobelman92}:
\begin{eqnarray}
{\hat H} &=& \sum\limits_{j=1}^{N} \sum\limits_{m=-1}^{1}
\hbar \left( \omega_0 - m^2 \delta \right) | e_{jm} \rangle
\langle e_{jm}|
\nonumber \\
&+&
\sum\limits_{\bm{\epsilon} \perp \mathbf{k}} \hbar ck
\left( {\hat a}_{\mathbf{k} \bm{\epsilon}}^{\dagger} {\hat a}_{\mathbf{k}\bm{\epsilon}} + \frac12 \right)
- \sum\limits_{j=1}^{N} {\hat{\mathbf{D}}}_j \cdot {\hat{\mathbf{E}}}(\mathbf{r}_j)
\nonumber \\
&+& \frac{1}{2 \varepsilon_0}
\sum\limits_{j \ne n}^{N} {\hat{\mathbf{D}}}_j \cdot {\hat{\mathbf{D}}}_n \delta(\mathbf{r}_j - \mathbf{r}_n).
\label{ham}
\end{eqnarray}
Here $| e_{jm} \rangle$ denotes the excited state of an atom $j$ having a magnetic quantum number $m_e = m$,
$\omega_0$ is the frequency of the transition $\left| J_g = 0 \right> \to \left| J_e = 1, m_e = 0 \right>$ in the presence of the field, $\hbar\delta$ is the energy difference (a Stark shift) between the excited states with $m_e = 0$ and $m_e = \pm 1$ (see Fig.\ \ref{figlevels} for an energy diagram),
${\hat a}_{\mathbf{k} \bm{\epsilon}}^{\dagger}$ and ${\hat a}_{\mathbf{k}\bm{\epsilon}}$ are creation and annihilation operators corresponding to an electromagnetic mode with a wave vector $\mathbf{k}$ and a polarization $\bm{\epsilon}$,
${\hat{\mathbf{D}}}_j$ are atomic dipole operators,
$\varepsilon_0 {\hat{\mathbf{E}}}(\mathbf{r}_j)$ are electric displacement vectors at atomic positions $\vec{r}_j$.
Note that ${\hat{\mathbf{E}}}(\mathbf{r})$ in Eq.\ (\ref{ham}) does not include the static field which is taken into account via the Stark shift $\delta$.

A derivation analogous to those in the absence of external field \cite{sokolov11,skip14} and in the presence of an external magnetic field \cite{skip15,skip16pra,skip18} allows us to reduce Eq.\ (\ref{ham}) to an effective non-Hermitian Hamiltonian of the atomic subsystem given by a $3N \times 3N$ `Green's matrix' $G$ with elements \cite{sokolov17,sokolov18}
\begin{eqnarray}
G_{e_{j m} e_{n m'}} &=& \left(i + 2 m^2 \Delta \right) \delta_{e_{j m} e_{n m'}} +
\frac{2 k_0^3}{\hbar \Gamma_0} (1 - \delta_{e_{j m} e_{n m'}})
\nonumber \\
&\times&
\sum\limits_{\mu, \nu}
{d}_{e_{j m} g_j}^{\mu} {d}_{g_n e_{n m'}}^{\nu}
\frac{e^{i k_0 r_{jn}}}{k_0 r_{jn}}
\nonumber
\\
&\times& \left[
\vphantom{\frac{r_{jn}^{\mu} r_{jn}^{\nu}}{r_{jn}^2}}
 \delta_{\mu \nu}
P(i k_0 r_{jn})
+ \frac{r_{jn}^{\mu} r_{jn}^{\nu}}{r_{jn}^2}
Q(i k_0 r_{jn})
\right],
\label{green}
\end{eqnarray}
where $P(x) = 1 - 1/x + 1/x^2$, $Q(x) = -1 + 3/x - 3/x^2$,
$\Gamma_0$ is the decay rate of the excited stated of an isolated atom, $\vec{r}_{jn} = \vec{r}_j - \vec{r}_n$, $\vec{d}_{e_{j m} g_j} = \langle J_{e} m|{\hat{\mathbf{D}}}_j | J_{g} 0 \rangle$, and $\Delta = \delta/\Gamma_0$ is the dimensionless Stark shift.
\red{
Values of $\Delta \sim 1$ are typical for existing experiments with clouds of cold atoms in external electric fields for field intensities of the order of several kV/cm \cite{snigirev14}. However, much stronger fields (up to 400 kV/cm) were previously used to study Stark effect in hot atomic beams \cite{krenn97}. Applying strong fields to cold atoms should, in principle, allow reaching much larger Stark shifts $\Delta \sim 100$ or even more.
}

The use of effective Hamiltonians is common in the theory of open quantum systems \cite{fyodorov97,dittes00,alhassid00} and implies that the eigenvalues $\Lambda_{\alpha}$ and right eigenvectors $\bm{\Psi}_{\alpha}$ of the effective Hamiltonian $G$,
\begin{eqnarray}
G {\bm \Psi}_{\alpha} = \Lambda_{\alpha} {\bm \Psi}_{\alpha},
\label{evequation}
\end{eqnarray}
play a role that is similar to the role played by the eigenvectors and eigenvalues of a standard Hermitian Hamiltonian in a closed system. Namely, any state ${\bm \Psi}$ of our open atomic subsystem can be represented as a superposition of eigenvectors (or `quasimodes') ${\bm \Psi}_{\alpha}$:
\begin{eqnarray}
{\bm \Psi} = \sum\limits_{\alpha} A_{\alpha} {\bm \Psi}_{\alpha},
\label{state}
\end{eqnarray}
whereas the eigenfrequencies $\omega_{\alpha} = \omega_0 - (\Gamma_0/2) \mathrm{Re} \Lambda_{\alpha}$ and decay rates $\Gamma_{\alpha}/2 = (\Gamma_0/2) \mathrm{Im} \Lambda_{\alpha}$ of quasimodes are determined by the complex eigenvalues $\Lambda_{\alpha}$. In the time domain, a short initial excitation will produce ${\bm \Psi}(t)$ equal to a weighted sum of oscillating and decaying terms $\propto \exp(-i \omega_{\alpha}t - \Gamma_{\alpha}t/2)$ with weights proportional to the spatial overlap of the initial excitations with ${\bm \Psi}_{\alpha}$.

The name `Green's matrix' for the effective Hamiltonian $G$ reflects the fact that each off-diagonal element of the matrix defined by Eq.\ (\ref{green}) is given by the Green's function of Maxwell equations describing the propagation of a monochromatic electromagnetic wave from a point source at $\mathbf{r}_n$ to a point $\mathbf{r}_j$ in the free space. Resonant scattering and Anderson localization of other types of waves (e.g., scalar waves \cite{skip16prb} or elastic waves \cite{skip18prb}) can be studied in the same framework by replacing the expression for the off-diagonal elements of the matrix $G$ by the Green's function of the corresponding wave equations. It is worthwhile to note that the use of Green's matrices to study the multiple scattering of waves has a long history starting with the papers by Foldy \cite{foldy45} and Lax \cite{lax51} and including the early work on Anderson localization of scalar waves \cite{rusek00,pinheiro04} and light \cite{rusek96}. The approach has been recently extended to aperiodic media \cite{wang18,sgr18}.

\section{Eigenvalues and eigenvectors of the Green's matrix}
\label{evalandevec}

\begin{figure*}
\hspace*{-1cm}
\includegraphics[width=1.1\textwidth]{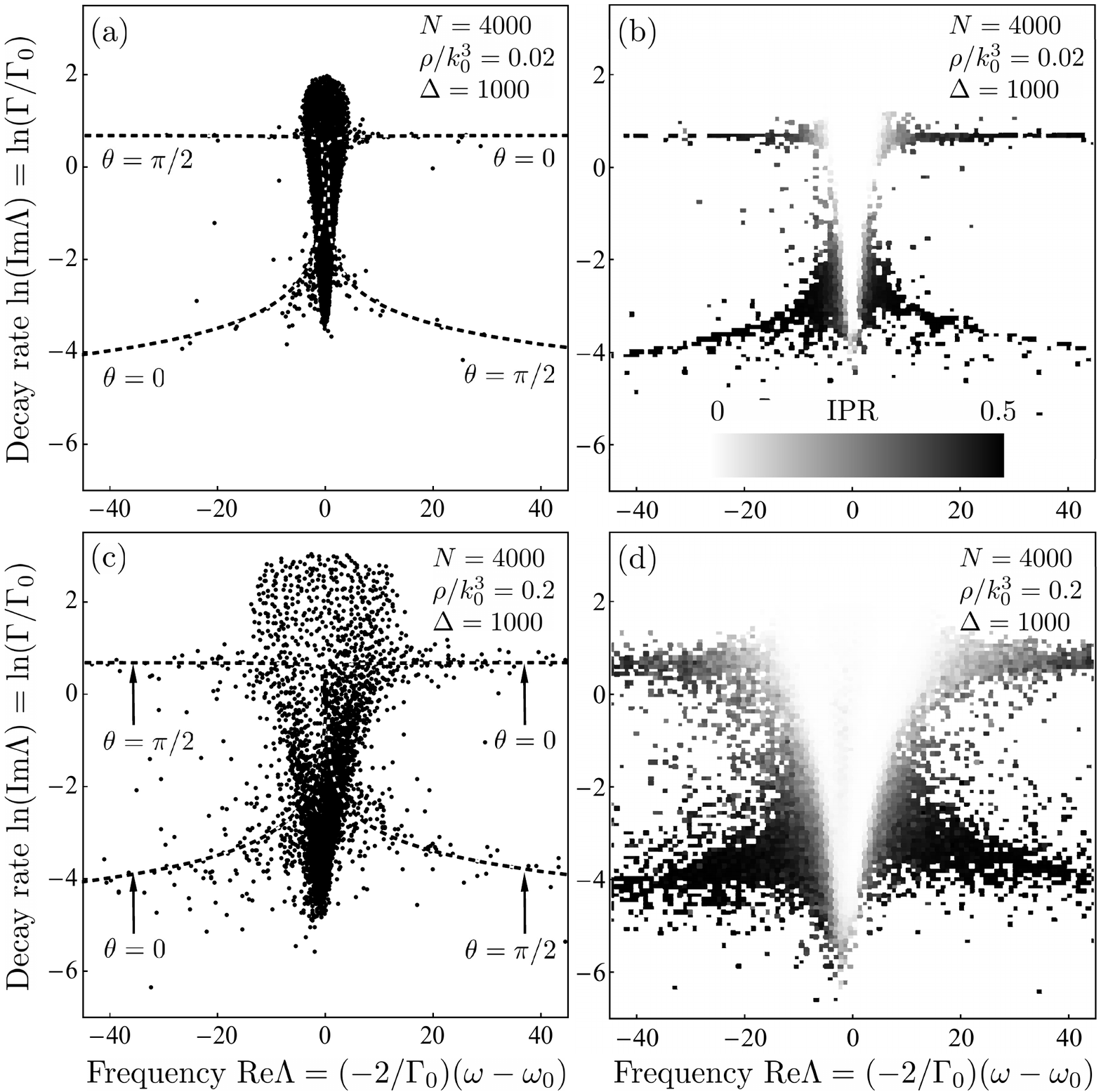}
\caption{Eigenvalues $\Lambda$ of the Green's matrix $G$ with $\mathrm{Re} \Lambda$ around 0 corresponding to probing the transition $\left| J_g = 0 \right> \to \left|J_e = 1, m_e = 0 \right>$, for a representative random configuration of $N = 4000$ atoms in a sphere
of radius $R$
at low (a) and high (c) densities.
\red{Eigenvalues of a pair of atoms would be located along one of the two pairs of the dashed lines depending on the angle $\theta$ between the line connecting the atoms and the external electric field, see Eq.\ (\ref{ev12}).}
Panels (b) and (d) show greyscale plots of the average IPR of eigenvectors at the same two densities (averaging is performed over 23 independent random configurations of atoms). The Stark shift is $\Delta = 1000$ for all four panels.
}
\label{figcenter}
\end{figure*}

\begin{figure*}
\hspace*{-1cm}
\includegraphics[width=1.1\textwidth]{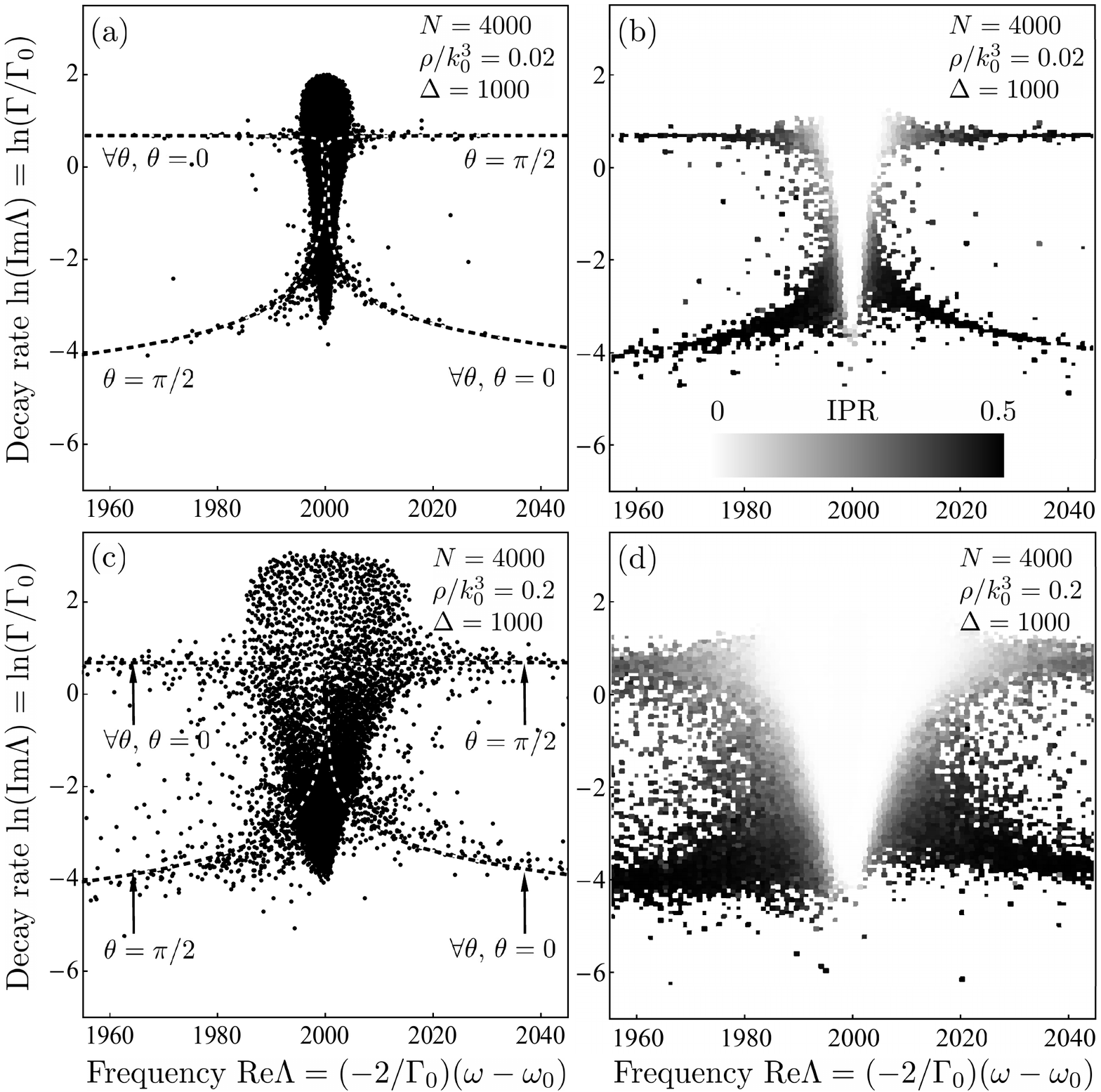}
\caption{Same as Fig.\ \ref{figcenter} but for the eigenvalues $\Lambda$ with $\mathrm{Re} \Lambda$ around $2 \Delta$ corresponding to probing the transition $\left| J_g = 0 \right> \to \left|J_e = 1, m_e = \pm 1 \right>$.
\red{Eigenvalues located along the dashed lines denoted by ``$\forall \theta$, $\theta = 0$'' correspond to Eq.\ (\ref{ev34}) for any angle $\theta$ or to Eq.\ (\ref{ev56}) for $\theta = 0$.}
}
\label{figright}
\end{figure*}

We study the statistical properties of eigenvalues $\Lambda_{\alpha}$ and eigenvectors $\bm{\Psi}_{\alpha}$ of the matrix $G$ by generating random configurations of scatterers $\left\{ \textbf{r}_j \right\}$ and finding $\Lambda_{\alpha}$ and $\bm{\Psi}_{\alpha}$ for each configuration numerically using a computer. When the external electric field is weak, $\Delta \lesssim 1$, the properties of eigenvalues and eigenvectors are similar to those found in the absence of the field \cite{skip14}. For strong fields $\Delta \to \infty$, the eigenvalues split in two well-separated groups corresponding to the single-atom transitions $|J_g = 0 \rangle \to |J_e = 1, m_e = 0 \rangle$ and $|J_g = 0 \rangle \to |J_e = 1, m_e = \pm 1 \rangle$, respectively. The eigenvalues corresponding to the first transition are concentrated around a line $\mathrm{Re} \Lambda = 0$ on the complex plane [see Fig.\ \ref{figcenter}(a) and (c)] whereas the eigenvalues corresponding to the second transitions have $\mathrm{Re} \Lambda \simeq 2 \Delta$ [see Fig.\ \ref{figright}(a) and (c)]. This is similar to the situation encountered in the presence of an external magnetic field \cite{skip15,skip18}, except that the transition $|J_g = 0 \rangle \to |J_e = 1, m_e = \pm 1 \rangle$ remains two-fold degenerate and the number of eigenvalues with $\mathrm{Re} \Lambda \simeq 2 \Delta$ is twice as large as the number of eigenvalues with $\mathrm{Re} \Lambda$ around zero.

By analogy with previous studies of scalar waves \cite{skip16prb}, light \cite{skip14,skip15,skip18} and elastic waves \cite{skip18prb}, we expect the eigenvectors of $G$ to be extended at low number densities $\rho$ of atoms
($\rho = N/V$, where $V = 4\pi R^3/3$ is the volume in which the atoms are distributed).
This is indeed confirmed by the calculation of the inverse participation ratio (IPR) of eigenvectors which quantifies the degree of eigenvector localization:
\begin{eqnarray}
\mathrm{IPR}_{\alpha} = \sum\limits_{j = 1}^{N} \left[ \sum\limits_{m = -1}^{1}
\left| \Psi_{\alpha j m} \right|^2 \right]^2,
\label{ipr}
\end{eqnarray}
where $\Psi_{\alpha j m}$ denotes the $m$-th component of the eigenvector ${\bm \Psi}_{\alpha}$ on the atom $j$ and we assumed that the eigenvectors are normalized:
$\sum_{j = 1}^{N} \sum_{m = -1}^{1} | \Psi_{\alpha j m} |^2 = 1$. Figures \ref{figcenter}(b) and \ref{figright}(b) show greyscale density plots of the average IPR as a function of the corresponding eigenvalue $\Lambda$. They have to be compared with the eigenvalue plots in the panels (a) of Figs. \ref{figcenter} and \ref{figright}. We see that the regions of the complex plane in which most of the eigenvalues are concentrated correspond to low values of IPR whereas IPR becomes significant only in the `branches' of the eigenvalue distribution, where very few eigenvalues are found for a given atomic configuration. The latter branches correspond to eigenvectors localized on pairs of closely located atoms and have been previously shown to exist for all types of waves \cite{skip14,skip15,skip16prb,skip18prb}.
\red{Analytic expressions of these branches can be readily obtained, see Eqs.\ (\ref{ev12})--(\ref{ev56}) in Sec.\ \ref{disc}. They are shown by dashed lines in Figs.\ \ref{figcenter}(a,c) and \ref{figright}(a,c).}
The properties of \red{eigenvectors belonging to the branches} do not evolve with the size of the atomic cloud and thus they do not obey the scaling expected for Anderson-localized states \cite{skip16prb}. We thus conclude that at a low number density of atoms $\rho$, the eigenvectors of the Green's matrix are extended.

Our previous results for atoms in a strong magnetic field suggest that localized eigenvectors may be expected to appear for large densities $\rho$ \cite{skip15}. However, this is not what we find from our calculation, see Figs.\ \ref{figcenter}(c,d) and \ref{figright}(c,d). Clearly, the groups of eigenvalues corresponding to the transitions $|J_g = 0 \rangle \to |J_e = 1, m_e = 0 \rangle$ (Fig.\ \ref{figcenter}) and $|J_g = 0 \rangle \to |J_e = 1, m_e = \pm 1 \rangle$ (Fig.\ \ref{figright}) widen with increasing $\rho$ but neither the imaginary parts of the eigenvalues decrease significantly (which would correspond to the appearance of long-lived states) nor the IPR of the eigenvectors increases in the central parts of eigenvalue groups where most of the eigenvalues are concentrated. We recall that an increase of density from $\rho/k_0^3 = 0.02$ to $\rho/k_0^3 = 0.2$ for atoms in a magnetic field causing the same frequency shift $\Delta$ as in Figs.\ \ref{figcenter} and \ref{figright}, leads to a decrease of the minimum value of $\mathrm{Im} \Lambda$ by several orders of magnitude accompanied by a growth of IPR for the eigenvectors corresponding to the eigenvalues in a narrow band of frequencies near $\mathrm{Re} \Lambda \simeq \pm 2 \Delta$ \cite{skip15}. These eigenvectors were shown to exhibit Anderson localization by the further analysis \cite{skip15,skip18}. In contrast, a strong electric field does not seem to produce the same effect and no signature of Anderson localization is seen in the evolution of eigenvalues and eigenvectors of the Green's matrix upon increasing the number density of atoms.

\section{Scaling analysis}
\label{scaling}

\begin{figure*}
\hspace*{-1cm}
\includegraphics[width=1.2\columnwidth]{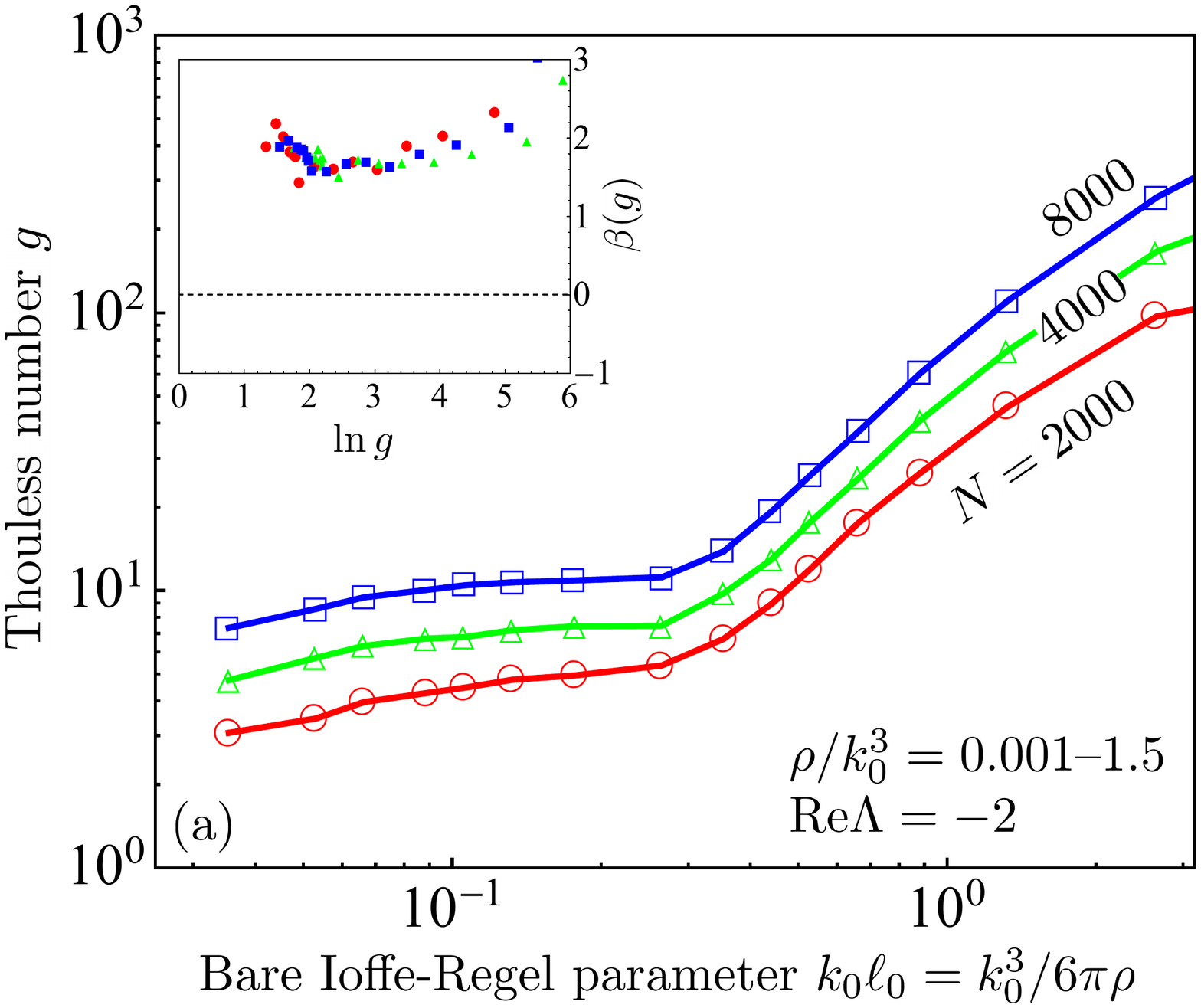}
\hspace*{-3cm}
\includegraphics[width=1.2\columnwidth]{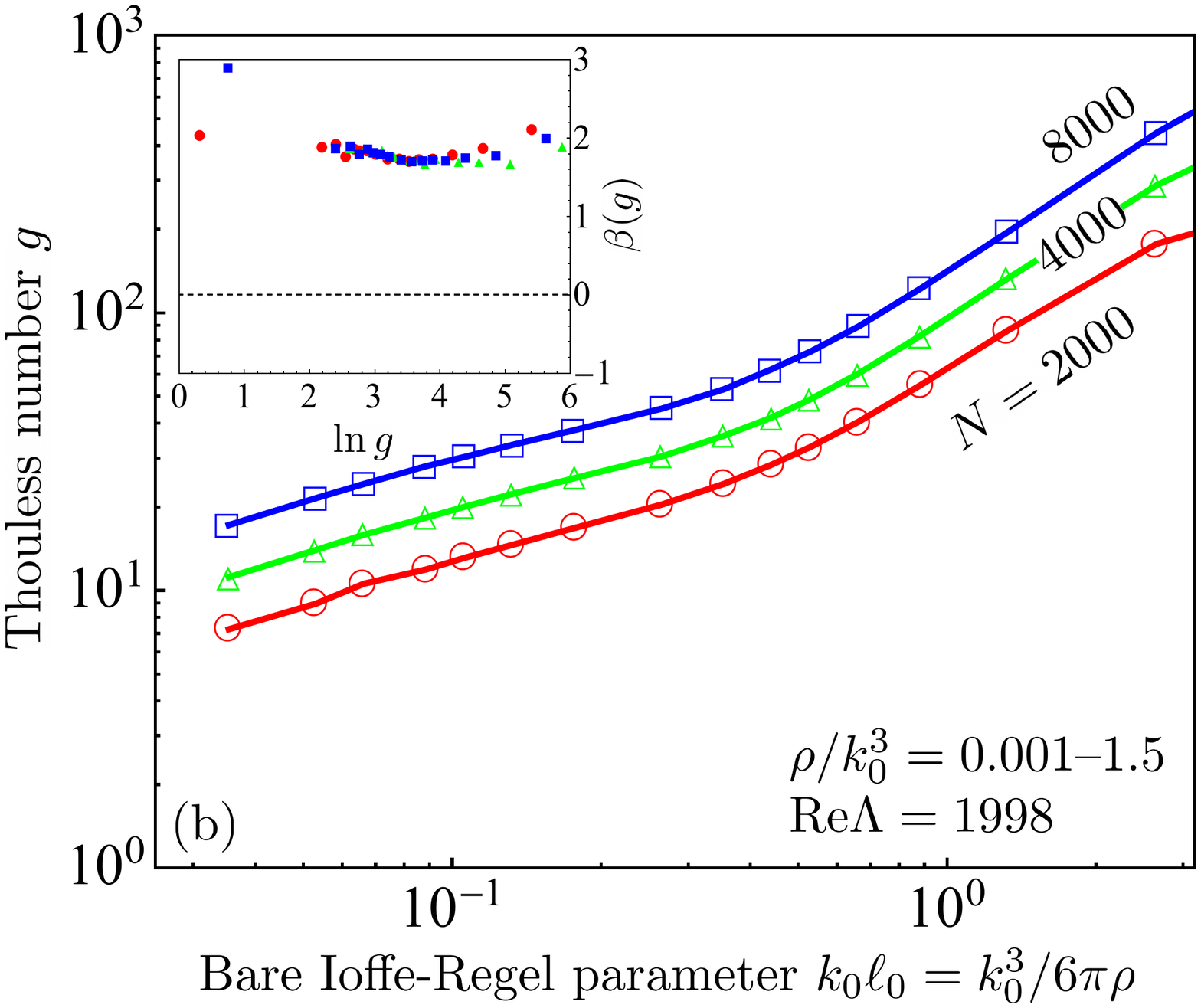}
\vspace*{-1.2cm}
\caption{Thouless number $g$ as a function of the bare Ioffe-Regel parameter $k_0\ell_0$ for three different numbers of atoms $N = 2000$, 4000 and 8000 (curves from bottom to top). $g$ is determined by averaging over eigenvalues in a unit interval of $\mathrm{Re} \Lambda$ around $\mathrm{Re} \Lambda = -2$ (a) or $1998$ (b), corresponding to probing the transitions $\left| J_g = 0 \right> \to \left|J_e = 1, m_e = 0 \right>$ (a) or $\left| J_g = 0 \right> \to \left|J_e = 1, m_e = \pm 1 \right>$ (b), respectively. The insets show the $\beta$-function $\beta(g) = \partial \ln g/\partial \ln k_0 R$ estimated by a finite-difference approximation of the derivative from the results corresponding to $N = 2000$ and 4000 (red circles), $N = 4000$ and 8000 (green triangles), and $N = 2000$ and 8000 (blue squares).}
\label{figscaling}
\end{figure*}

The analysis of the previous section is quite pictorial but qualitative and does not exclude that Anderson localization effects may strengthen and eventually become dominant when the size $R$ of the considered atomic system is increased at a fixed atomic number density $\rho$. Indeed, the scaling of relevant physical quantities with the size of a disordered system proved to be a reliable indicator of localization properties of eigenstates \cite{abrahams79,pichard81,mackinnon83}. If localization effects grow with $R$, and provided that one works with $R$ that are already sufficiently large (e.g., larger than all other relevant length scales, such as the wavelength or the mean free path), one may conclude by extrapolation that they will dominate in the thermodynamic limit $R \to \infty$. In contrast, if localization effects decrease with $R$, it is likely that they will become negligible for $R \to \infty$. Here we will apply this idea to the Thouless number \cite{wang11,skip14}
\begin{eqnarray}
g(\mathrm{Re} \Lambda) = \frac{\langle (\mathrm{Im} \Lambda_{\alpha})^{-1} \rangle_{\mathrm{Re} \Lambda_{\alpha} \in B}^{-1}}{\langle \mathrm{Re} \Lambda_{\alpha+1} -  \mathrm{Re} \Lambda_{\alpha} \rangle_{\mathrm{Re} \Lambda_{\alpha} \in B}},
\label{thouless}
\end{eqnarray}
where the averaging $\langle \cdots \rangle_{\mathrm{Re} \Lambda_{\alpha} \in B}$ is performed over eigenvalues in a narrow band $B = [\mathrm{Re} \Lambda - B/2, \mathrm{Re} \Lambda + B/2]$ of frequencies $\mathrm{Re} \Lambda_{\alpha}$. We stress that it is important to analyze $g$ with frequency resolution instead of averaging over all eigenvalues, which would correspond to $B \to \infty$. This is due to the strong frequency dependence of the properties of the resonant atomic system that we consider. Indeed, strong scattering and potentially Anderson localization of light may be expected only near the atomic resonance (in our case, for $\omega \simeq \omega_0$ and $\omega \simeq \omega_0 - \delta$, corresponding to $\mathrm{Re} \Lambda \simeq 0$ and $\mathrm{Re} \Lambda \simeq 2 \Delta$, respectively), whereas the interaction of light with atoms is weak far from the resonances. Averaging over all frequencies would mix up different types of behavior and may yield uncontrollable results.

Thouless number $g$ is a ratio between the typical decay rate of quasimodes in the numerator and the average mode spacing in the denominator of Eq.\ (\ref{thouless}).
\red{
Without averaging in the numerator of Eq.\ (\ref{thouless}), $g$ would be a random quantity with a very wide probability distribution near the mobility edge and under conditions of Anderson localization \cite{shapiro86,shapiro87}. Quantitative analysis of Anderson localization may require analysis of its full probability distribution \cite{slevin2001,marcos2006}. A practical way to perform such an analysis has been proposed by Slevin \textit{et al.} \cite{slevin2003} and applied to the model of wave scattering by point scatterers in recent works \cite{skip16prb,skip18prb,skip18}. In contrast to the latter works, we are not aiming at rigorous quantitative results here, so that
}
the simplified definition (\ref{thouless}) will be sufficient for our purposes.

Figure\ \ref{figscaling} shows $g$ as a function of a bare Ioffe-Regel parameter $k_0 \ell_0 = k_0^3/6\pi\rho$ for two frequency intervals around $\mathrm{Re} \Lambda = 0$ (a) and $\mathrm{Re} \Lambda = 2 \Delta$ (b), respectively, and for three different numbers of atoms $N$ [and hence for three different radii $R = (3 N/4 \pi \rho)^{1/3}$ of the atomic system at each density $\rho$]. Similar behavior is observed for other frequencies. The choice of $k_0 \ell_0$, with $\ell_0$ the on-resonance scattering mean free path in the independent-scattering approximation (ISA), as a control parameter is convenient because the localization transition takes place at $k_0 \ell_0 \simeq 1$ for scalar waves \cite{skip18ir}. Therefore, the largest number density of atoms $\rho/k_0^3 = 1.5$ represented in Fig.\ \ref{figscaling} exceeds the density required to reach Anderson localization of scalar waves by more than an order of magnitude whereas the smallest $\rho/k_0^3 = 0.001$ corresponds to a very dilute medium. However, no sign of Anderson localization is seen in Fig.\ \ref{figscaling} despite the wide range of explored densities. Thouless number $g$ increases with $N$ (and hence with $R$) at a given $\rho$ for all $\rho$, which is manifest in the fact that the three curves in Fig.\ \ref{figscaling}(a) and (b) are roughly parallel to each other and do not show any tendency to cross. A crossing between curves $g(k_0 \ell_0)$ corresponding to different $N$ would indicate a change in behavior and a possible Anderson transition \cite{skip14,skip15} but it is not observed in Fig.\ \ref{figscaling}.

A more rigorous way of expressing the fact that Fig.\ \ref{figscaling} does not show any sign of Anderson transition consists in computing the so-called $\beta$-function defined as \cite{abrahams79}:
\begin{eqnarray}
\beta(g) = \frac{\partial \ln g}{\partial \red{\ln} k_0 R}.
\label{beta}
\end{eqnarray}
From the asymptotic behavior of $g$ for localized and extended modes, we readily conclude that $\beta < 0$ in the first case, $\beta > 0$ in the second case, and $\beta = 0$ at the critical point of the localization transition. Another important message of Eq.\ (\ref{beta}) is the so-called single-parameter scaling---$\beta$ is assumed to depend on a single parameter $g$ and not on $N$, $\rho$, $R$, etc. separately. We show $\beta(g)$ estimated from the numerical data of Fig.\ \ref{figscaling}(a) and (b) in the insets of corresponding panels. We see that indeed $\beta(g)$ always remains positive, signaling the absence of localized states in our model, and that all points obtained by combining the data corresponding to different $N$ and $\rho$ roughly fall on a single master curve. This confirms the absence of Anderson localization in the considered model.

\section{Discussion}
\label{disc}

In order to understand the reasons behind so different impacts of static magnetic and electric fields on the phenomenon of Anderson localization of light by atoms, we diagonalize the matrix (2) for $N = 2$ atoms separated by a distance $r$ and take the limit of strong field $\Delta \to\infty$. The six eigenvalues reduce to
\begin{eqnarray}
\Lambda_{1,2} &=& \red{i} \pm \frac32 \frac{e^{i k_0 r}}{k_0 r}
\left[ P(i k_0 r) + Q(i k_0 r) \cos^2 \theta \right],
\label{ev12}
\\
\Lambda_{3,4} &=& 2 \Delta \red{+ i} \pm \frac32 \frac{e^{i k_0 r}}{k_0 r} P(i k_0 r),
\label{ev34}
\\
\Lambda_{5,6} &=& 2 \Delta \red{+ i} \pm \frac32 \frac{e^{i k_0 r}}{k_0 r}
\left[ P(i k_0 r) + Q(i k_0 r) \sin^2 \theta \right],\;\;\;\;\;\;
\label{ev56}
\end{eqnarray}
where $\theta$ is the angle between the vector $\mathbf{r}$ and the external electric field.

The first two eigenvalues given by Eq.\ (\ref{ev12}) correspond to the transition $|J_g = 0 \rangle \to |J_e = 1, m_e = 0 \rangle$ and coincide with those found in the presence of a strong magnetic field \cite{skip15}. The properties of this transition are therefore the same and it does not lead to Anderson localization of light as we discussed previously \cite{skip15}. The four remaining eigenvalues correspond to the transition $|J_g = 0 \rangle \to |J_e = 1, m_e = \pm 1 \rangle$ which remains two-fold degenerate. For this reason and in contrast to the case of magnetic field, we cannot introduce an effective scalar model for this transition even in the limit of $\Delta \to \infty$. What we can do, however, is to compare Eqs.\ (\ref{ev34}) and (\ref{ev56}) with the corresponding results in the absence of any field and in the presence of the magnetic field. Such a comparison may help us to point out similarities and difference between the three cases, although it will not allow us to make unambiguous conclusions with certainty.

First, the two eigenvalues $\Lambda_{3,4}$ given by Eq.\ (\ref{ev34}) coincide with those of the Green's matrix (\ref{green}) in the absence of external fields ($\Delta = 0$), except for the trivial shift of $2 \Delta$. It is also worthwhile to mention that each of these eigenvalues would be two-fold degenerate in the absence of fields. The eigenvalues $\Lambda_{3,4}$ can thus be seen as originating from the physical processes that also take place in the absence of external fields and that, as we know already \cite{skip14}, do not lead to Anderson localization.

Second, the eigenvalues $\Lambda_{5,6}$ given by Eq.\ (\ref{ev56}) are similar to those found in the presence of an external magnetic field but the latter contain an additional factor $\frac12$ in front of $Q(i k_0 r)$. As a result, the magnitude of the near-field part of $\Lambda_{5,6}$ (i.e., of the part that diverges faster that $1/r$ for $r \to 0$ and contains terms describing the dipole-dipole interaction between the atoms) is a factor of 2 larger than the magnitude of the equivalent term for atoms in an external magnetic field.
\red{We see therefore that an external electric field appears to be less efficient than the magnetic field in suppressing near-field, dipole-dipole coupling between neighboring atoms.}

\begin{table*}[t]
\caption{\label{tabcomp}
\red{
Coefficients in front of far-field ($\alpha_{\mathrm{far}}$) and near-field ($\alpha_{\mathrm{near}}$) parts of eigenvalues of a two-scatterer system for different models.}
}
\begin{ruledtabular}
\begin{tabular}{llllccl}
Wave                     &             & $\alpha_{\mathrm{far}}(\theta)$ & $\alpha_{\mathrm{near}}(\theta)$  &  $\langle |\alpha_{\mathrm{near}}(\theta)|^2 \rangle_{\theta}$  & Localization & Ref.\\
\colrule
Scalar                   &             & 1          & $\hphantom{-}0$           & 0 &  Yes & \onlinecite{skip14}, \onlinecite{skip16prb}\\
\colrule
Electromagnetic          &             & $\frac32$  & $\hphantom{-}\frac32$ & $\frac94$ &  \multirow{2}{*}{No} & \multirow{2}{*}{\onlinecite{skip14}}  \\
without external fields  &             & 0          & $-3$      & 9 &                    &  \\
\colrule
Electromagnetic          & $m = 0$     & $\frac32(1-\cos^2\theta)$ & $\hphantom{-}\frac32(1-3\cos^2\theta)$ & $\frac95$ & No   & \multirow{2}{*}{\onlinecite{skip15}, \onlinecite{skip18}}                 \\
with magnetic field      & $m = \pm 1$ & $\frac34(1+\cos^2\theta)$ &  $-\frac34(1-3\cos^2\theta)$ & $\frac{9}{20}$ & Yes &                   \\
\vspace{-3.5mm}\\
\colrule
Electromagnetic          & $m = 0$     & $\frac32(1-\cos^2\theta)$ & $\hphantom{-}\frac32(1-3\cos^2\theta)$ & $\frac95$ & &     This             \\
with electric field      & $m = \pm 1$ & $\frac32$ & $\hphantom{-}\frac32$ & $\frac94$ & No  &  work                \\
                         & $m = \pm 1$ & $\frac32\cos^2\theta$ & $-3(1-\frac32\cos^2\theta)$ & $\frac{81}{20}$ &  &                  \\
\end{tabular}
\end{ruledtabular}
\end{table*}

\red{
All the available results concerning Anderson localization in 3D ensembles of resonant point scatterers \cite{skip14,bellando14,skip15,skip18,skip16prb,skip18prb} support the idea, first launched in Ref.\ \cite{skip14}, that the possibility to realize Anderson localization is correlated with the strength of near-field (or dipole-dipole) interactions between pairs of scatterers. In particular, for scalar and electromagnetic waves (with and without external fields), we can write the eigenvalues $\Lambda$ of the Green's matrix of a two-atom system in a universal form:
\begin{eqnarray}
\Lambda &=& \Lambda_0 \pm \frac{e^{i k_0 r}}{k_0 r}
\left\{ \vphantom{\left[ \frac{i}{k_0 r} - \frac{1}{(k_0 r)^2} \right]}
\alpha_{\mathrm{far}}(\theta)
\right. \nonumber \\
&+& \left. \alpha_{\mathrm{near}}(\theta)
\left[ \frac{i}{k_0 r} - \frac{1}{(k_0 r)^2} \right]
\right\},
\label{evuni}
\end{eqnarray}
where the first term $\Lambda_0 = \mathrm{Re}{\Lambda_0} + i$ with $\mathrm{Re}{\Lambda_0}$ representing an irrelevant frequency shift, the coefficient $\alpha_{\mathrm{far}}$ gives the magnitude of the far-field coupling between scatterers, whereas the coefficient $\alpha_{\mathrm{near}}$ measures the strength on near-field effects. Both $\alpha_{\mathrm{far}}$ and $\alpha_{\mathrm{near}}$ may, in general, depend on the angle $\theta$ between the vector $\vec{r}$ connecting the scatterers and the direction of an external magnetic or electric field, if present. For elastic waves, $\Lambda$ of a two-scatterer system cannot be recast into Eq.\ (\ref{evuni}) because two different wave numbers $k_p$ and $k_s$ appear for pressure (longitudinal) and shear (transverse) waves instead of a single wave number $k_0$ in Eq.\ (\ref{evuni}). However, the divergence of $\Lambda$ for $r \to 0$ is the same as for scalar waves (i.e., $\Lambda \propto 1/r$), leading to a similar behavior as far as Anderson localization is concerned \cite{skip18prb}.
}

\red{
We summarize our results for $\alpha_{\mathrm{far}}$ and $\alpha_{\mathrm{near}}$ in Table\ \ref{tabcomp}. Different models differ by the absolute values of these complex coefficients, their phase and anisotropy. However, a perfect correlation with the existence of Anderson localization in a given model can be established only for the magnitude of near-field interactions between atoms that we characterize by an angle-averaged square of the absolute value of $\alpha_{\mathrm{near}}(\theta)$:
\begin{eqnarray}
\langle |\alpha_{\mathrm{near}}(\theta)|^2 \rangle_{\theta} =
\frac{1}{2} \int\limits_0^{\pi} |\alpha_{\mathrm{near}}(\theta)|^2 \sin(\theta) d\theta.
\label{angleav}
\end{eqnarray}
This quantity is also given in Table\ \ref{tabcomp}.
Our results suggest that $\langle |\alpha_{\mathrm{near}}(\theta)|^2 \rangle_{\theta}$ should be smaller than a critical value of order 1 (more precisely, a value between 9/20 = 0.45 and 9/5 = 1.8)  for Anderson localization to exist. This conjecture should be taken with great care because it is based on the analysis of only a small number of discrete values of $\langle |\alpha_{\mathrm{near}}(\theta)|^2 \rangle_{\theta}$ corresponding to physically realizable situations. In order to prove the validity of this conjecture, it would be necessary to consider a physically sound model in which the strength of near-field interactions could be varied continuously.
}

\section{Conclusion}
\label{concl}

The crucial role of an external magnetic field for reaching Anderson localization of light in a random arrangement of immobile atoms \cite{skip15,skip18} may suggest that the same or similar effect may be achieved by other physical mechanisms that lift the degeneracy of atomic states. In this work we explore one of such mechanisms---the Stark effect---by considering collective quasimodes of large ensembles of randomly distributed, immobile atoms in a static external electric field. We study the spatial structure of the quasimodes and characterize their localization in space by the inverse participation ratio (IPR). In addition, we compute the Thouless number $g$ equal to the ratio of the typical decay rate of quasimodes to the average frequency spacing between quasimodes, and analyze the evolution of $g$ with the size of the atomic cloud at a fixed atomic number density. Our main conclusion is that the electric field does not induce Anderson localization of light by the atoms. The reason for this is the fact that, in contrast to the magnetic field, the electric field lifts the degeneracy of atomic states only partially. As a result, the three-fold degenerate excited state with the total angular moment $J_e = 1$ splits in a nondegenerate state $| J_e=1, m_e=0 \rangle$ and a two-fold degenerate state $| J_e=1, m_e=\pm 1 \rangle$. The transition between a nondegenerate ground state $|J_g = 0\rangle$ and $| J_e=1, m_e=0 \rangle$ has exactly the same properties as the corresponding transition in a strong magnetic field studied previously \cite{skip15}. It does not lead to Anderson localization of light that is quasiresonant with it. The transition between $|J_g = 0\rangle$ and $| J_e=1, m_e=\pm 1 \rangle$ is different from the corresponding transitions in the magnetic field. Our analysis shows that the strength of the near-field, dipole-dipole coupling between atoms due to the exchange of photons with frequencies that are quasiresonant with this transition, is smaller than in the absence of external fields but larger than in the presence of a magnetic field. As a result, the light that is quasi-resonant with this transition does not exhibit Anderson localization either.

\red{
A joint analysis of results obtained in this work together with the previously published results concerning Anderson localization of scalar waves and light by resonant point scatterers suggests that the strength of near-field, dipole-dipole interactions between scatterers should be less than a certain critical value for Anderson localization to take place. This conjecture calls for a verification in a model where the strength on near-field interactions could be varied continuously.
}

\begin{acknowledgments}
This work was funded by the Agence Nationale de la Recherche (Grant No. ANR-14-CE26-0032 LOVE).
Part of the work concerning the analysis of the spectrum of collective atomic states was carried out with the financial support of the Russian Science Foundation (Project No. 17-12-01085). IMS acknowledges the hospitality of the LPMMC where a part of this work has been performed with a financial support of the Centre de Physique Th\'{e}orique de Grenoble-Alpes (CPTGA).
\end{acknowledgments}

\end{document}